\documentstyle[12pt,epsbox,psfig]{article}

\advance \textheight by \topskip
\begin{document}

\title{Multi-Megaton Water Cherenkov Detector\\ 
for a Proton Decay Search\\
--- TITAND (former name: TITANIC) ---}

\author{Y. Suzuki\\
Kamioka Observatory, ICRR, University of Tokyo\\
Higashi-Mozumi, Yoshiki-gun, Kamioka-cho\
Gifu, 506-1205 Japan\\
(for the TITAND Working Group\protect\cite{TitanicMember})} 

\maketitle


\section{Introduction}

The discovery of the neutrino oscillations in 1998 revisited the 
interest of proton decay search~\cite{Pd-theory}. 
	The existence of the tiny neutrino mass indicates the 
existence of the large energy scale behind and suggests the connection 
between the small neutrino mass and proton decay. 
	There are strong indications of the unification scale of 
O(10$^{16}$)GeV: the running coupling constants with super-symmetric 
particles merge at this large energy scale. 
	Proton decay directly brings us to this unification scale.

The search for proton decay has been conducted for more than 30 years 
and currently the most stringent limits are set mostly by 
the Super-Kamiokande experiment~\cite{SK-Pdecay}. 
	The limit for e$^{+}\pi^0$ mode
is now 5.3$\times$10$^{33}$ years (for 84ktyr), and 
that for $\bar{\nu}$K$^{+}$ mode is 1.9$\times$10$^{33}$ years. 
	The recent theoretical development by lattice QCD suggests 
that all the 
predictions on the life time of proton should be shortened from the 
previous calculations about 
an order of magnitudes~\cite{Pdecay_QCD} and therefore the search 
for proton decay may become much more realistic than before. 

The current interested region of the proton decay life time is in the 
range beyond about 10$^{35}$ years for p $\rightarrow$ e$^{+}\pi^{0}$ 
mode and 10$^{29-35}$ yrs for p $\rightarrow$ K$^{+}\bar{\nu}$ 
mode~\cite{Life-Pdecay}.

In this report, we will show a detector
which can be used to search for proton decay in the lifetime region 
beyond 10$^{35}$ years. 
	We will briefly review the current experimental status and 
discuss the sensitivity of the future proton decay detectors, and we 
specifically present 
a possibility of a scalable multi-megaton water Cherenkov detector 
immersed in the shallow water.

\section{Current Status of the Proton Decay Life Time Measurements}

Most stringent limits for proton decay comes from the Super-Kamiokande 
experiment, the 50,000tons imaging water Cherenkov detector, located
1000m underground.
	For the details of the Super-Kamiokande experiment, see
references~\cite{SK-Pdecay}.
	We explain the current status of the search for the typical 
decay modes of p $\rightarrow$ e$^{+}\pi^{0}$ and 
p $\rightarrow$ K$^{+}\bar{\nu}$.
	The 1289 days of data which corresponds to 79.3kty were used 
for the searches.
 	
\subsection{p $\rightarrow$ e$^{+}\pi^{0}$}

The event candidates are selected by applying the following selection 
criteria: 1) there are 2 or 3 Cherenkov Rings; 2) all rings are 
showering; 3) the reconstructed $\pi^{0}$ mass should be between 85 
and 185 MeV/c$^{2}$ for 3 Ring events; 4) there are no electrons 
from muon decay; 5) proton mass range 
should be between 800 and 1050 MeV/c$^{2}$; and 6) the total momentum 
should be balanced to be less than 250 MeV/c. 
	The efficiency for this selection, 43\%, is 
estimated by the Monte Carlo calculation and 
the backgrounds from
atmospheric neutrino interactions are estimated to be 0.2 events. 
	We found no candidates, and then set the limits on the partial 
life of 5.0$\times$10$^{33}$years for 
p$\rightarrow$e$^{+}\pi^{0}$(90\% C.L.).   

\subsection{p$\rightarrow$K$^{+}\bar{\nu}$}

We have adopted three different searches for the 
p$\rightarrow$K$^{+}\bar{\nu}$ mode. 
	Since the momentum of K$^{+}$ from the proton decay 
is less than 
the Cherenkov threshold, and therefore the K$^{+}$ is invisible 
in the detector. 
	The decay products of K$^{+}$ must be detected to 
identify the decay modes. 
For the case of K$^{+}\rightarrow\mu^{+}\nu$, the 
momentum of $\mu$ is 236 MeV/c, and
there are large backgrounds from
atmospheric neutrino interactions producing single visible muon ring. 

Additional information may be obtained from the prompt $\gamma$-ray 
emission by the de-excitation of the $^{15}$N after the 
disappearance of proton in the nuclei by the proton decay. 
	For about 50\% of the case, they emit 6.4 MeV 
$\gamma$-rays~\cite{nu-K-gamma}. 
The gammas can be identified before the muon event time. 
No candidate events with the prompt $\gamma$-rays were found. 

The third method is to look for K$^{+}\rightarrow\pi^{+}\pi^{0}$ mode.
	Those $\pi^{0}$ must be reconstructed and the momentum flow 
of the two gammas may be restricted in the momentum range between 
180 and 250 MeV/c, where the expected momentum is 205 MeV/c. 
	The $\pi^{+}$ gives some scattered
light opposite to the $\pi^{0}$ direction. 
The requirement of this scattered
lights may separate the signal events from the backgrounds. 
	We found one candidate event where 
about two backgrounds are expected. 

By combining those three methods, 
we have obtained the limit of 
$\tau$/B(p$\rightarrow$K$^{+}$$\bar{\nu}$)
=1.6$\times$10$^{33}$years.

\section{Sensitivity of the Future Proton Decay Experiment}

First we consider the p$\rightarrow$e$^{+}$$\pi^{0}$ mode.
	If we simple apply the same selection criteria that 
Super-Kamiokande is using, 45 background events can be found in the
signal box for the 20 Mton$\cdot$yr of data by the MC simulation.
	This background rate corresponds to about 2.2 events per 
Mton$\cdot$year. 
Based on this background estimate, the sensitivity of the water 
Cherenkov 
detector, a la Super-Kamiokande, is 1$\times$10$^{35}$years
for 10 years of 1 Mton (fiducial) detector and 
5$\times$10$^{35}$years for 
10 years of 10 Mton (fiducial) detector (with 90\% CL sensitivities). 

In the following discussion we assume that we use the same PMTs 
SK is using, namely
50cm diameter PMTs.
	For a large detector, we may need to reduce the density of the 
number of photo-multiplier tubes. 
	We have studied the case for the reduced density of
1/2 and 1/3 of the SK density. 
	The reduction of the detection efficiency for those reduced 
density is about 70\% and 50\% for the number density of 
1/2 and 1/3, respectively.

The sensitivity of the current analysis, 
is limited by the level of backgrounds as was seen in the discussion
above. 
	An improved analysis is possible by applying a tight cut on the 
total momentum. 
	The tight 100 MeV/c cut (instead of 250MeV/c) looses efficiency 
down to 17.4\%, but the most of the signal comes from the decay 
of free protons. 
	Therefore there are no Fermi momentum, no binning energy and 
no nuclear effect corrections. 
	Then the systematic error of the detection efficiency would 
become smaller than the previous analysis.
	The background rate for this tight cut is 3 events per 
20 Mton$\cdot$yr ($\sim$0.15 events/Mton$\cdot$yr). 
	The 3$\sigma$ discovery limit with 
this tight cut reaches to $\sim$4$\times$10$^{35}$ years for 10 years of
10 Mton detector, and $\sim$7$\times$10$^{34}$ years for 10 years of 1 Mton
detector. 
UNO and HyperK (with $\sim$500kton fiducial volume) reaches to 
$\sim$3$\times$10$^{34}$ years.

\begin{figure}
\begin{center}
%\rule{5cm}{0.2mm}\hfill\rule{5cm}{0.2mm}
%\vskip 2.5cm
%\rule{5cm}{0.2mm}\hfill\rule{5cm}{0.2mm}
%\epsbox{file=pd.eps,height=7cm}
\psfig{file=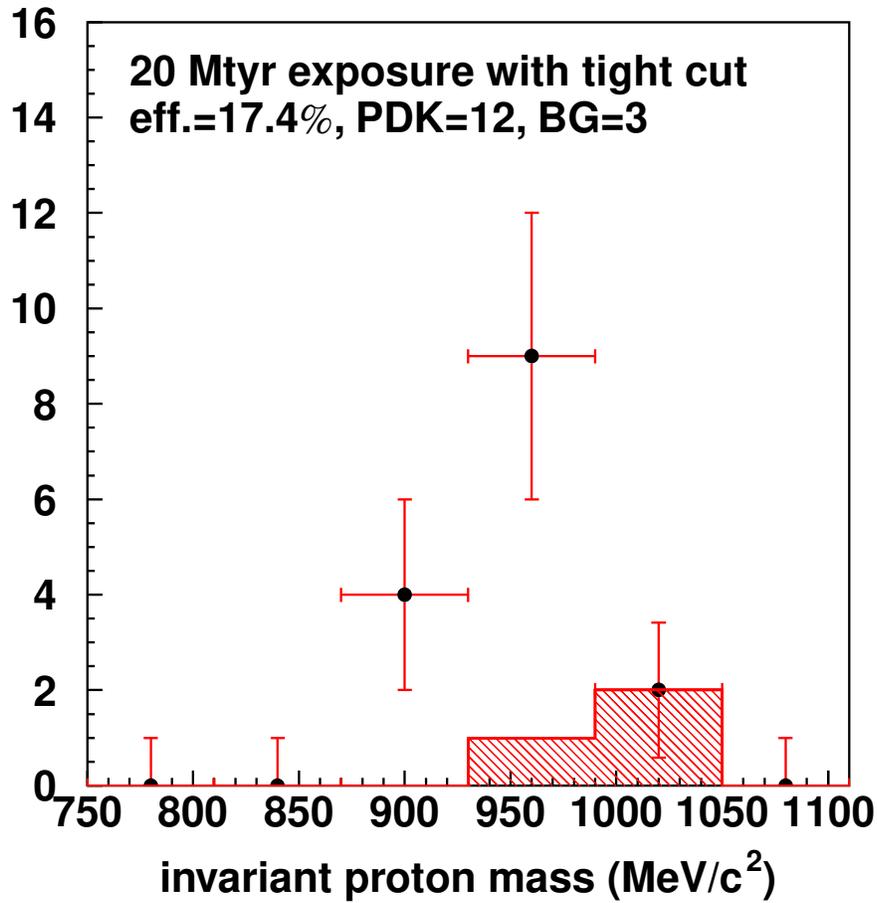,height=12cm}
\caption{Proton decay peak. The 12 signal events are cleanly seen over
the 3 background events from atmospheric neutrino
interactions.}
\label{fig:inv_mass}
\end{center}
\end{figure}

Fig~\ref{fig:inv_mass}. shows the invariant mass distribution for the
case of the proton life time is 1$\times$10$^{35}$. 
	You can clearly see the peak from proton decay. 
The signal to backgrounds(S/N) is 4 and 1 for 1$\times$10$^{35}$ and
4$\times$10$^{35}$ years life time, respectively. 
	Proton decay searches 
into the life time longer than 10$^{35}$years are become possible only 
by the detector heavier than a few Mega-tons.
%%%%%%
%%%%

Backgrounds for the $\nu$K$^{+}$ mode has also been estimated. 
	The most effective channel for the search is to look for 
the prompt $\gamma$'s from the nuclear de-excitation of $^{15}$N. 
We expect 6 event per Mega$\cdot$ton$\cdot$year from such backgrounds.
	The background for the single muon search and 
$\pi^{+}\pi^{0}$ search are 2100 and 22 events per Mega$\cdot$ton$\cdot$year, 
respectively. 

\begin{figure}[t]
\begin{center}
%\rule{5cm}{0.2mm}\hfill\rule{5cm}{0.2mm}
%\vskip 2.5cm
%\rule{5cm}{0.2mm}\hfill\rule{5cm}{0.2mm}
%\epsbox{file=pd.eps,height=7cm}
\psfig{file=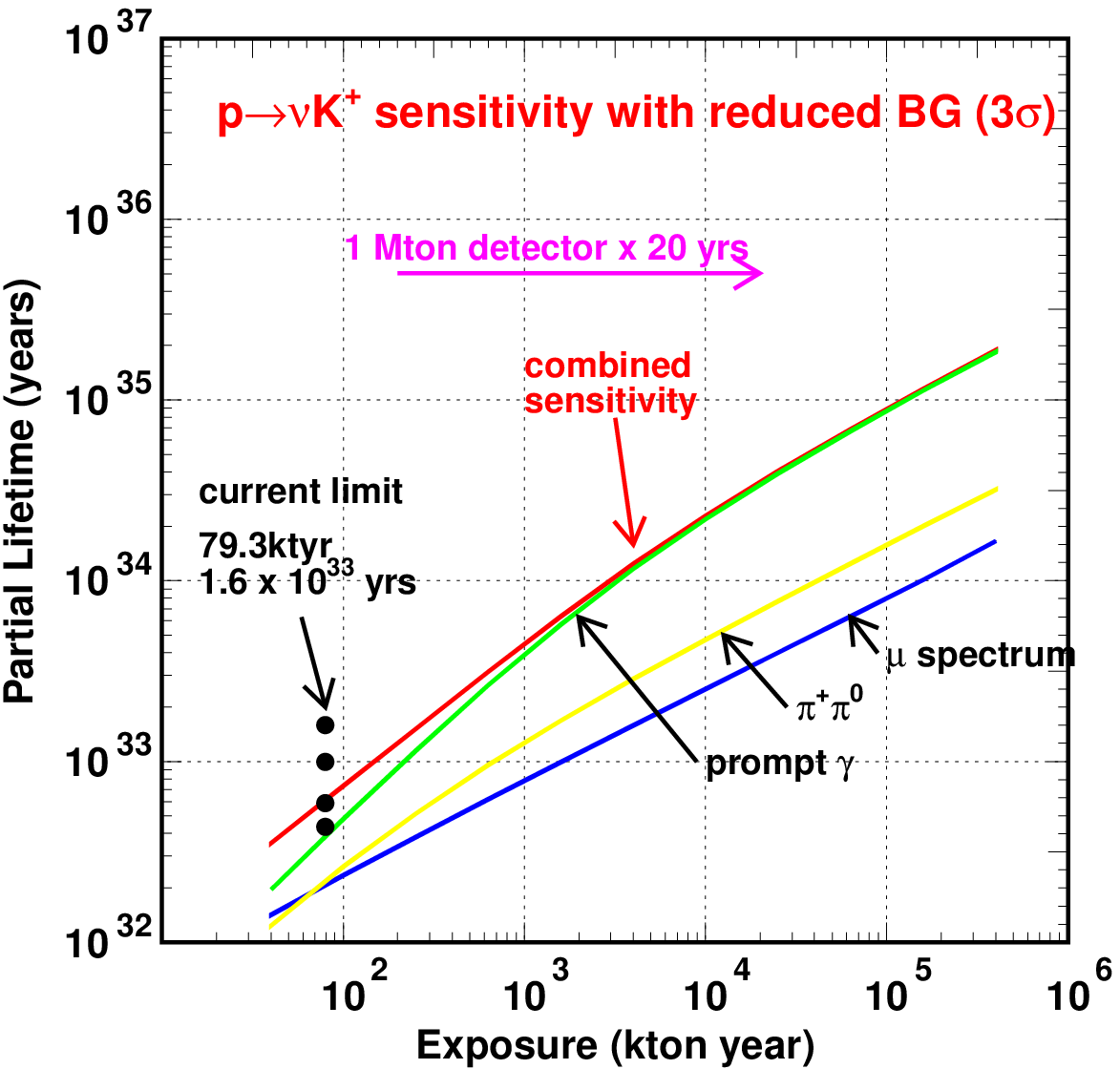,height=12cm}
\caption{The sensitivity of the $\bar{\nu}$K$^{+}$mode.
The most sensitive mode is the single muons from K decay with the $\sim$6 MeV
prompt $\gamma$'s.
For a 20 Mtonyr exposures, you can reach to the life time sensitivity of 
6$\times$10$^{34}$years.}
\label{fig:nu-K_sensitivity}
\end{center}
\end{figure}

The K production by the atmospheric neutrinos, expected to be about 
1 events per Mega$\cdot$ton$\cdot$year, 
becomes very serious background, which would
also give a prompt gamma ray. The life time sensitivity for the proton 
decay of such mode is shown in Fig.~\ref{fig:nu-K_sensitivity}.
%%%%%%
%%%%

\section{Underground vs Underwater}

There are proposals to build a Mega-ton scale underground water Cherenkov
detector. 
	The fiducial volumes of those detectors are around 0.5 Mtons.
The volume is limited by the maximum side 
we can build the cavity in the underground with current technology.
	Underground detectors may also have a difficulty to expand 
their physical size, after the construction has been completed.
	The length of the construction (cavity excavation) is another 
limitations: it is estimated that it takes about 7 years to build 
such cavities. 
	The cost also one of the limiting factors.
	Of course, they can be built by a simple application of 
the current technology, by which we have build Super-Kamiokande.

	As was stated in the previous section, the discovery limits 
of the 1Mt (fiducial volume) detector is 7$\times$10$^{34}$ in ten 
years.
	Therefore if one want to reach the lifetime region of 10$^{35}$
years, then we need other approaches.

\begin{figure}[t]
\begin{center}
%\rule{5cm}{0.2mm}\hfill\rule{5cm}{0.2mm}
%\vskip 2.5cm
%\rule{5cm}{0.2mm}\hfill\rule{5cm}{0.2mm}
%\epsbox{file=pd.eps,height=7cm}
\psfig{file=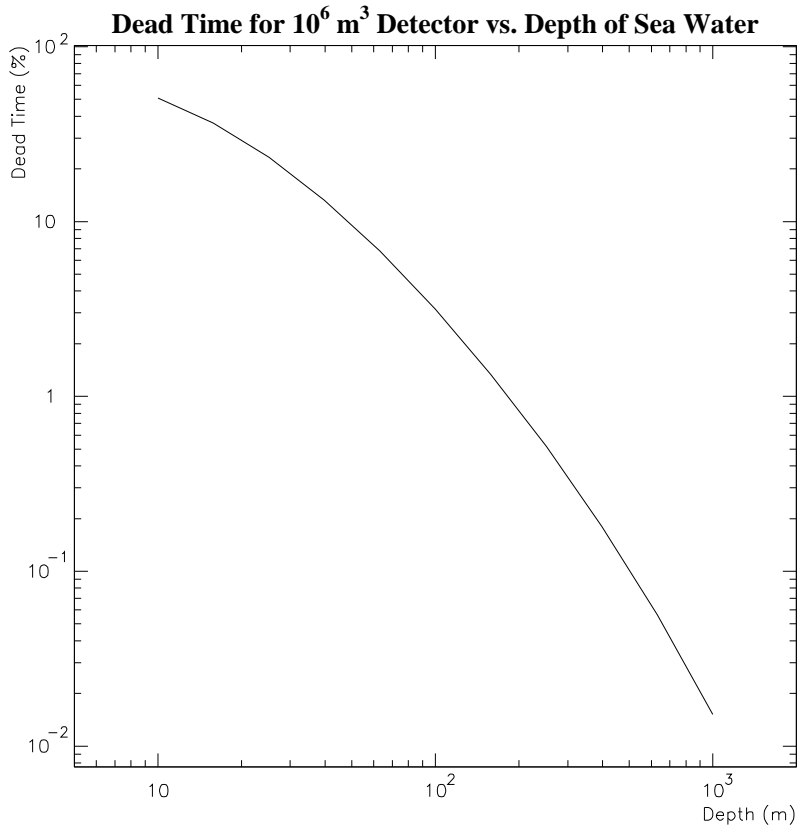,height=12cm}
\caption{Dead time created by the cosmic ray muons as a function of 
the depth of the detector.
For 1Mton detector, you can deduce the background to be less than 3\%
if you can go deeper than 100m. We assumed that 1$\mu$sec dead time is
created for each muons passing through the detector.}
\label{fig:murate}
\end{center}
\end{figure}

Here we consider the shallow under water detectors, placed about 100 m
under the surface of the water. 
	This 100 m requirement comes from the necessity
to reduce the muon rate. 
	Suppose that each muon passing through the detector produces
1 $\mu$sec dead time corresponding to the time for muons passing 
through the detector, then as shown in Fig. $\ref{fig:murate}$, 
100 m depth is necessary to keep the dead time to be less than 3\%.
%%%%%%
%%%%%%
Such detector has some advantages over underground Mega-ton detectors:
it has an expand-ability and it can be built in short construction
time of about 2 to 3 years. 
The construction cost is lower than the underground detectors.

Disadvantage is the followings.
	In such detector, solar neutrino measurements are not possible 
due to the numerous cosmic muons producing the spallation backgrounds. 
	For $\bar{\nu}$K search, we may require the precise time and 
spatial coincidence to identify the prompt $\sim$6MeV $\gamma$-rays over those 
backgrounds.

\section{The Multi-Megaton Water Cherenkov Detector---TITAND}

\begin{figure}[t]
\begin{center}
%\rule{5cm}{0.2mm}\hfill\rule{5cm}{0.2mm}
%\vskip -3cm
%\rule{5cm}{0.2mm}\hfill\rule{5cm}{0.2mm}
%\epsbox{file=pd.eps,height=7cm}
%\psfig{file=detector1.eps,height=10cm,width=7cm}
\psfig{file=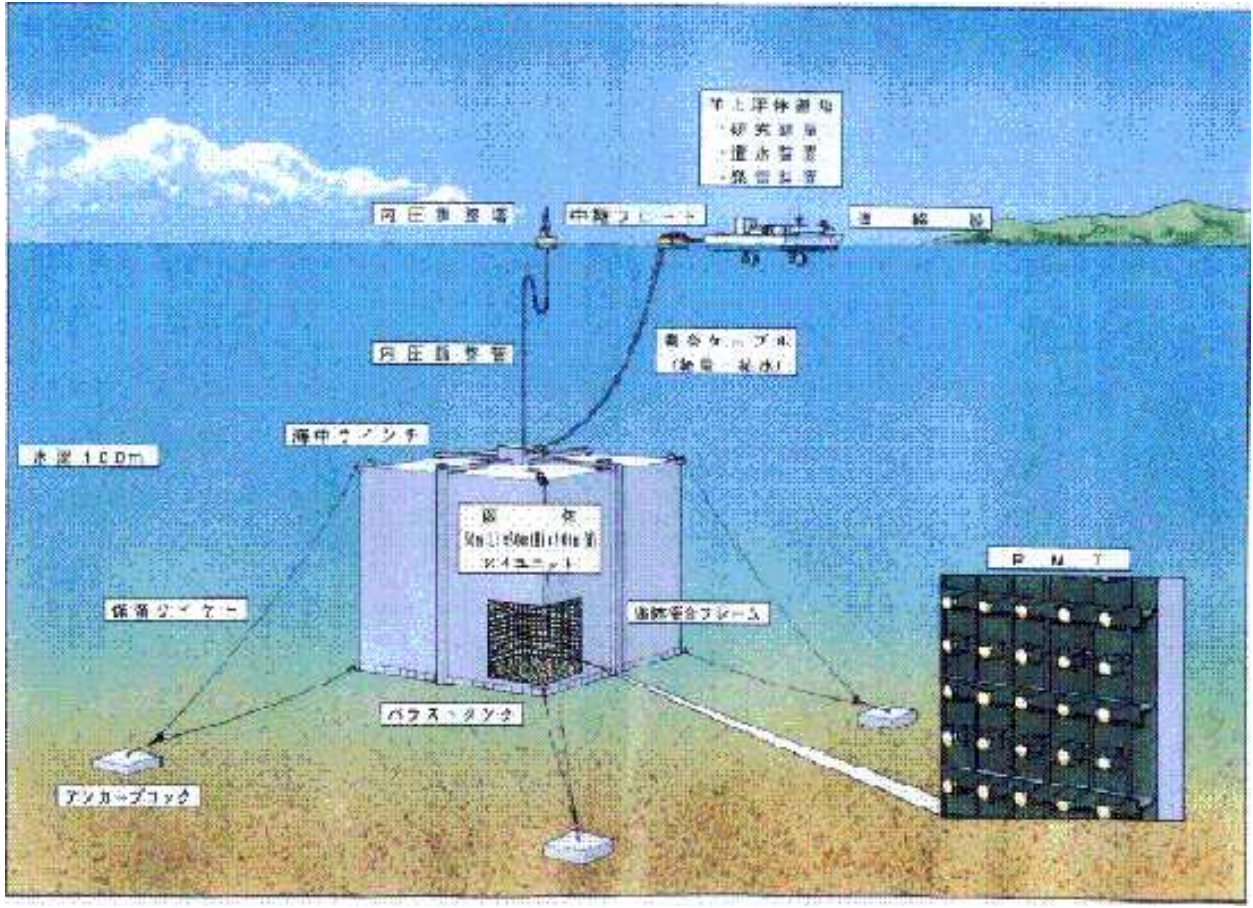,height=12cm}
\caption{The artist's view of the detector. The unit detector (2 Mton) consists
of 4 identical modules assembled at the experimental site. About 60K
PMTs are used for 2Mt detector which provides a 20\% photo-coverage
of the inner surface.}
\end{center}
\label{fig:detector}
\end{figure}

\begin{figure}[t]
\begin{center}
%\rule{5cm}{0.2mm}\hfill\rule{5cm}{0.2mm}
%\vskip 2.5cm
%\rule{5cm}{0.2mm}\hfill\rule{5cm}{0.2mm}
%\epsbox{file=pd.eps,height=7cm}
\psfig{file=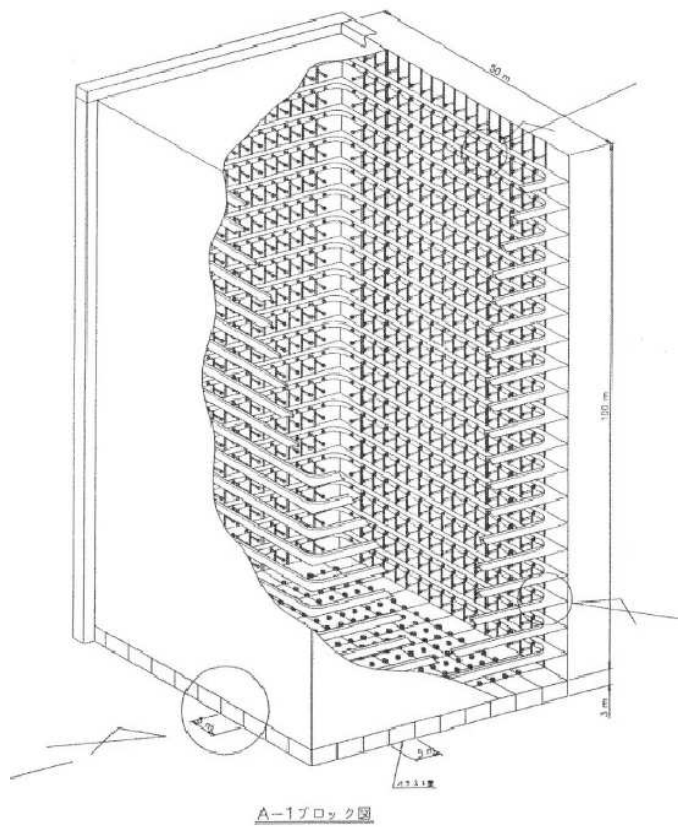,height=12cm}
\caption{The mechanical structure of the module.The frame structure 
is also used for mounting PMTs and the container is made as a
semi-pressure vessel which hold the pressure difference of about 1MPa.}
\label{fig:structure}
\end{center}
\end{figure}

The detector called TITAND(Totally Immersible Tank Assaying Nucleon 
Decay) consists of 4 identical modules with the 
dimension of 70mx70mx100m (Fig.~\ref{fig:detector}).
%%%%%%
%%%%%%
	The detector 140mx140mx100m may be the first phase of 
TITAND with 2 Mton water inside, and this can be expanded in future. 
	Inside of the detector is covered by PMT modules of 
50 cm diameter, arranged every 1m apart, which provide 20\% 
photo-coverage, same as Kamiokande, yet to be a good resolution 
detector for relatively higher energy event detection. 
	The outer part of the detector layer with 2m thickness on an 
average acts as a active-shield to identify incoming particles, 
especially cosmic ray muons.
	Four modules are put together at the experimental site 
to make the basic unit.
	The vessel is a semi-pressure type and hold the pressure 
difference upto 20 atmospheric pressure. 
	The frame structure, as shown in Fig.~\ref{fig:structure}, 
holds the vessel against the outside pressure. 
	The inner water level is kept to be 
10 atmospheric pressure at maximum, namely the water level inside is 
hold at the level of top of the water vessels, not at the sea level.
	Therefore the PMT does not need to be placed in a pressure 
container.
%%%%%%%%%%%%%%
%%%%%%%%%%%%%%
	There is another option to make the water level at the pressure of 
the sea level.
	For this option, the vessel can be made thinner material, 
however, the PMTs must hold against the high pressure, 
therefore they must be placed in the pressure container. 
This options are also under considerations.

The detector may not necessarily to be placed at the bottom of the sea.
	The density difference between the salt water and pure water
(3\%) is compensated the weight of the steel vessels. Then this detector
balances to the buoyancy forces and float at any depth of the sea.
	Actually the designed detector here have ballast tanks: we can
make the detector float; by removing the water from them and we can 
the detector sink by filling the water in them.

\section{Construction of the Detector}

The major construction can be made in the dock on the shore. 
	The maximum size of the dock
available in Japan is 80mx80mx400m. 
	Therefore the four 70mx70mx100m modules can be made
in the dock at the same time and all the internal structure, 
including the pluming, 
PMT mounting and so on can be done as well.
	The 4 modules made in the dock are towed by a ship to the 
experimental site, and start to fill the water made through the 
water desalinamation system and the water purification system.
	Those facility will be placed on the floating bulge, on which 
the generator, water purification system, research facility and 
dormitory will also be placed.
	The floating bulge is a sub-merged structure and equipped 
with the self positioning system and can be kept in the aimed place 
within an accuracy of 10 m.
	The ability of the water desalination system is 100 ton per 
hour and two sets can be used.

Term of the construction is relatively fast. 
	After one or two year of the design stage. 
	The actual construction of the Vessels can be finished in 
two years. 
	Detector construction time is, therefore limited by the time to 
make the light sensors. 
	But as usual case, once we have decided to make it, 
those equipments can be built within a desired time by, for example,
making new factories to make the required 
number of light sensors.

Cost of the detector for the proton decay experiments are crucial, 
since some may cost more than \$500k for those to be built 
underground.
	The construction cost should be substantially reduced.
	The current estimate of TITAND including all the components 
is \$250k for 2.0 M ton detector.

\section{Other Physics Opportunities}

TITAND will be used not only for searching for proton decay, 
but also used for the observation of neutrino burst from Supernovae,
a movable detector as a far detector for neutrino factories.

	 The signature of the 
supernova bursts is characterize as a very short pulsed events 
within 10 to 20 seconds and about half of the events occur within 
the first 0.5 seconds.
	Those characteristic signature can be measured relatively easily 
by TITAND.
	The sensitive region for the bursts is extended from those of
Super-Kamiokande to well cover those happening in Andromeda, 650kpc 
away from the earth. 
	We expect about 200 events/Mton for an Andromeda supernova.
	The detector can also detect atmospheric neutrinos and 
we expect about 100k events per year/Mton.
 
	TITAND is a movable detector and can even go deeper, 
therefore it is well suited for the far detector for the long 
baseline neutrino oscillation experiments either by using the 
super-beam or by neutrino factories.
	Other type of the detector can be placed as an additional 
module and can be attached to the proton decay module, therefore 
the ideal detector can be used for the neutrino oscillation 
experiments. 
	For example. magnetized ion detector could be added 
if necessary.

\end{document}